\documentclass[prb,twocolumn,showpacs,preprintnumbers,amsmath,amssymb, superscriptaddress]{revtex4-1}

\usepackage{amsmath}    
\usepackage{amssymb}
\usepackage{graphicx}   
\usepackage{verbatim}   
\usepackage{color}      
\usepackage{hyperref}   
\usepackage[normalem]{ulem}
\usepackage{natbib}
\usepackage{fixmath}
\usepackage{enumitem}
\usepackage{dsfont}

\def \brc #1{\left\lbrace #1 \right\rbrace}
\def \loc {\mathrm{loc}}

\hypersetup{colorlinks,linkcolor=blue,urlcolor=blue,citecolor=blue}
\newcommand{\beq}{\begin{equation}}
\newcommand{\eeq}{\end{equation}}
\newcommand{\bea}{\begin{eqnarray}}
\newcommand{\eea}{\end{eqnarray}}

\newcommand{\bbGamma}{{\mathpalette\makebbGamma\relax}}
\newcommand{\makebbGamma}[2]{%
  \raisebox{\depth}{\scalebox{1}[-1]{$\mathsurround=0pt#1\mathbb{L}$}}%
}

\newcommand{\be}{\begin{equation}}
\newcommand{\ee}{\end{equation}}

\definecolor{darkgreen}{rgb}{0,0.5,0}
\definecolor{orange}{rgb}{1,0.5,0}
\definecolor{grey}{rgb}{.6,.6,.6}

\newcommand{\tn}{\tilde n}

\newcommand{\tc}{\tilde c}

\newcommand{\bra}[1]{\langle #1|}
\newcommand{\ket}[1]{|#1\rangle}
\newcommand{\average}[1]{\langle #1\rangle}

\newcommand{\cL}{{\cal L }}
\newcommand{\fL}{\mathfrak{ L }}
\newcommand{\cH}{{\cal H }}
\newcommand{\fH}{\mathfrak{H }}

\newcommand{\ketL}[1]{|#1 )}
\newcommand{\braL}[1]{( #1|}

\newcommand{\tsigma}{{\tilde \sigma}}

\def\doubleunderline#1{\underline{\underline{#1}}}

\begin{document}
\title{Simulating Lindbladian evolution with non-abelian symmetries: Ballistic front propagation in the SU(2) Hubbard 
model with a localized loss}
 \author{C\u at\u alin Pa\c scu Moca}
\affiliation{Department of Theoretical Physics, Institute of Physics, Budapest University of Technology and Economics, Budafoki \'ut 8., H-1111 Budapest, Hungary}
\affiliation{Department of Physics, University of Oradea,  410087, Oradea, Romania}
\author{Mikl\'os Antal Werner}
\affiliation{Department of Theoretical Physics, Institute of Physics, Budapest University of Technology and Economics, Budafoki \'ut 8., H-1111 Budapest, Hungary}
\affiliation{MTA-BME Quantum Dynamics and Correlations Research Group, 
Institute of Physics, Budapest University of Technology and Economics,  Budafoki \'ut 8., H-1111 Budapest, Hungary}
\author{\" Ors Legeza}
\affiliation{Strongly Correlated Systems Lend\"ulet Research Group,
Wigner Research Centre for Physics, H-1525, Budapest, Hungary}
\affiliation{Institute for Advanced Study,Technical University of Munich, Lichtenbergstrasse 2a, 85748 Garching, Germany}
 \author{Toma\v z Prosen}
\affiliation{Department of Physics, Faculty of Mathematics and Physics,
University of Ljubljana, Jadranska 19, SI-1000 Ljubljana, Slovenia}	 
\author{M\'arton Kormos}
\affiliation{MTA-BME Quantum Dynamics and Correlations Research Group, 
Institute of Physics, Budapest University of Technology and Economics,  Budafoki \'ut 8., H-1111 Budapest, Hungary}
\author{Gergely Zar\'and}
\affiliation{Department of Theoretical Physics,  Institute of Physics, Budapest University of Technology and Economics,  Budafoki \'ut 8., H-1111 Budapest, Hungary}
\affiliation{MTA-BME Quantum Dynamics and Correlations Research Group, 
Institute of Physics, Budapest University of Technology and Economics,  Budafoki \'ut 8., H-1111 Budapest, Hungary}
\date{\today}
\begin{abstract}
We develop a  non-Abelian time evolving block decimation (NA-TEBD) approach 
to study of open systems governed by  Lindbladian time evolution, 
while exploiting  an arbitrary number of  abelian or non-abelian symmetries. 
We illustrate this method in a one-dimensional fermionic SU(2) Hubbard model  on a semi-infinite lattice
with localized particle loss at one end. We observe a ballistic front propagation 
with strongly renormalized front velocity, and a hydrodynamic current density profile. 
For large  loss rates, a suppression of   the particle current is observed, as a result  of  
the quantum Zeno effect. Operator entanglement is found to propagate faster than the depletion 
profile, preceding the latter.
\end{abstract}
\maketitle

\section{Introduction}

Understanding dynamical effects, correlations or entanglement properties in strongly correlated 
systems subject to dephasing or dissipation~\cite{Harbola.2008, Breuer.2016} or in systems under continuous monitoring~\cite{Chin.2010}
represents a major challenge for modern condensed matter physics~\cite{Muller.2012}.
%
Recent experimental advances in ultracold
atoms have made laboratory studies of the time dependent evolution of non-equilibrium states in such quantum many body systems possible~\cite{Diehl.2008, Schneider.2012,Chong.2018,Damanet.2019}. For example,
quantum quenching the interaction by means of Freschbach resonances, has triggered
enormous activity both experimentally~\cite{Syassen.2008,Cheneau.2012,Barontini.2013,Lapp.2019} and theoretically~\cite{Eckel.2010,Calabrese.2011,Bua.2012,Eisert.2015,Kormos.2017, Mitra.2018}. 
By now, state of the art experiments  
allow  the study of the dynamics of even  a single quantum level~\cite{Fukuhara.2013}, 
or the implementation of microscopic spin filters in quantum point contact cold atom setups~\cite{Lebrat.2019}.
In some of these high resolution experiments the fundamental effect of measurement backaction due to  an  external observer becomes significant,  and request 
a closer investigation of  open many-body systems,  where the interaction with the environment plays a major role.

A conceptually and mathematically consistent though computationally demanding  approach 
to model the external environment in  interacting systems  is the Lindblad approach~\cite{Wichterich.2007,Prosen.2012,Ajisaka.2012,Ajisaka.2012a,Pizorn.2013,Karevski.2013,Ajisaka.2014,Arrigoni.2013,Dorda.2014,Dorda.2015,Jin.2016,Schwarz.2016}.  
In the Lindbladian framework,  time evolution is described in terms of 
the density operator, $\rho$, and the linear, hermiticity-preserving map $\cal L$, the so-called Lindbladian,
 which generates the time evolution of  the density operator $\rho(t)$ via 
\bea
i \, \dot { \rho} &=&  {\cal L}[\, \rho\,] =    [ H, \rho]  + {\cal D}\,[\, \rho\,]
\,,  
\label{eq:Lindblad}
\eea
 with the dissipator map,  ${\cal D} = \sum_F \lambda_F \, {\cal D}_{F}$,  written as a convex combination of elementary maps
\bea
{\cal D}_F\,[\, \rho\,]  \equiv i\,  F\, \rho\,  F^\dagger -{\textstyle {i \over 2}}\, \big \{ \, F^{\dagger} F\,, \, \rho \,\big \}\,.
\eea 
The Lindblad equation  \eqref{eq:Lindblad}
describes  the most general   Markovian trace preserving  dynamics, 
and generates a completely positive non-unitary map  (a.k.a. quantum channel).
The so-called Lindblad jump operators $F$ shall be later on  simply referred to as {\em dissipators}.

Most of the numerical methods used to investigate  time evolution, such as the time-evolving block decimation 
(TEBD) ~\cite{Vidal.2003,Vidal.2004,Verstraete.2004, Zwolak.2004}, the time dependent density matrix renormalization 
group (TD-DMRG)\cite{Daley.2004,White.2004,Feiguin.2005} or time dependent variational principle
(TDVP)~\cite{Haegeman.2013, Zauner-Stauber.2018, Vanderstraeten.2019} have been originally designed for closed systems. 
They are rarely used for open systems, 
where they request  considerable computational effort and their use is therefore challenging~\cite{Schollwock.2005}.  
In fact, already for such simple systems  as the Hubbard chain, studied here, 
the dimension of the local  space  increases to $16$,  which is extremely difficult to handle with  standard matrix-product-state (MPS) methods. 
As we show here, efficient implementation of symmetries~\cite{Werner.2016,Hauschild.2018,Jaschke.2018,Weichselbaum.2020} and, in particular, 
\emph{non-Abelian symmetries}~\cite{Werner.2020}, makes it possible to efficiently simulate the dissipative time evolution of these 
models.

Symmetry operations in quantum mechanics are represented by certain unitary (or antiunitary) operators, $U$,  
which transform quantum states into other quantum states, $|\psi\rangle \to  |\psi_U \rangle \equiv U |\psi \rangle$. Similarly,  $U$ 
transforms the  density operator into another density operator, $\rho\to {\rho}_U= U\,\rho\,U^\dagger$. 
For closed systems, we call $U$ a \emph{symmetry} if it commutes with the Hamiltonian,  
$[U,H]=0$,  implying that time evolution commutes with the symmetry operation.
We can easily extend this concept to open systems, 
where it entails the condition\footnote{This type of symmetry has been defined in Ref.~\cite{Bua.2012} as {\em the weak symmetry}, in order to distinguish it from a more restrictive situation of a {\em strong symmetry} where $H$ and the full set of $\{F\}$ commute with $U$.}
\be 
U\, {\cal L}[\,\rho\,] \,U^\dagger =  {\cal L}[\,U\, \rho \,U^\dagger]\,. 
\label{eq:Lsymmetry}
\ee 
Eq.~\eqref{eq:Lsymmetry}  immediately implies that the time evolved density operator satisfies the relation, 
$\rho_U (t)  =   U \rho(t) U^\dagger$.  
As we show in Sections~\ref{sec:MPS} and \ref{sec:superfermion}, 
for non-Abelian symmetries,  Eq.~\eqref{eq:Lsymmetry} has certain 
implications regarding the  structure of the dissipator map: the groups of dissipators, $\{F_m\}$, related by symmetry transformations 
must occur with identical dissipation strength $\lambda_F$, otherwise they break the  
non-Abelian symmetry. 

In this paper we further develop the notion of symmetries in the Liouville space --- a vector space of density operators of the system, which allows us 
to  handle abelian and non-abelian symmetries in a transparent and efficient way.  Although non-abelian 
symmetries can be treated  within the matrix product operator (MPO) approach,\cite{Werner.future} here we follow
 a technically somewhat  simpler approach: we vectorize the density operator and 
 the Lindblad equation~\cite{Shallem.2015},  and represent the vectorized density matrix as an MPS.  We identify symmetry operations,  and apply  non-abelian MPS methods of Ref.~\onlinecite{Werner.2020} 
 in this augmented vector space. 

Some lattice models with local  losses 
have  been studied earlier.   Non-interacting fermions and  bosons, e.g.,  have been analyzed 
in Refs~\onlinecite{Krapivsky.2019} and \onlinecite{Sels.2020},  respectively.  Spinless fermions 
with nearest neighbor interactions~\cite{Froml.2019, Froml.2020,Wolff.2020} or in Bose-Hubbard models~\cite{Kepesidis.2012, Labouvie.2016}
have also been studied, however, spinful interacting models represent a major challenge due to the quickly increasing 
local operator space.  The approach presented here allows us to investigate \emph{spinful}
fermion and boson models with matrix product methods efficiently, in cases where some non-abelian symmetry is  
not broken by dissipation.

\begin{figure}[t!]
\begin{center}
 \includegraphics[width=0.95\columnwidth]{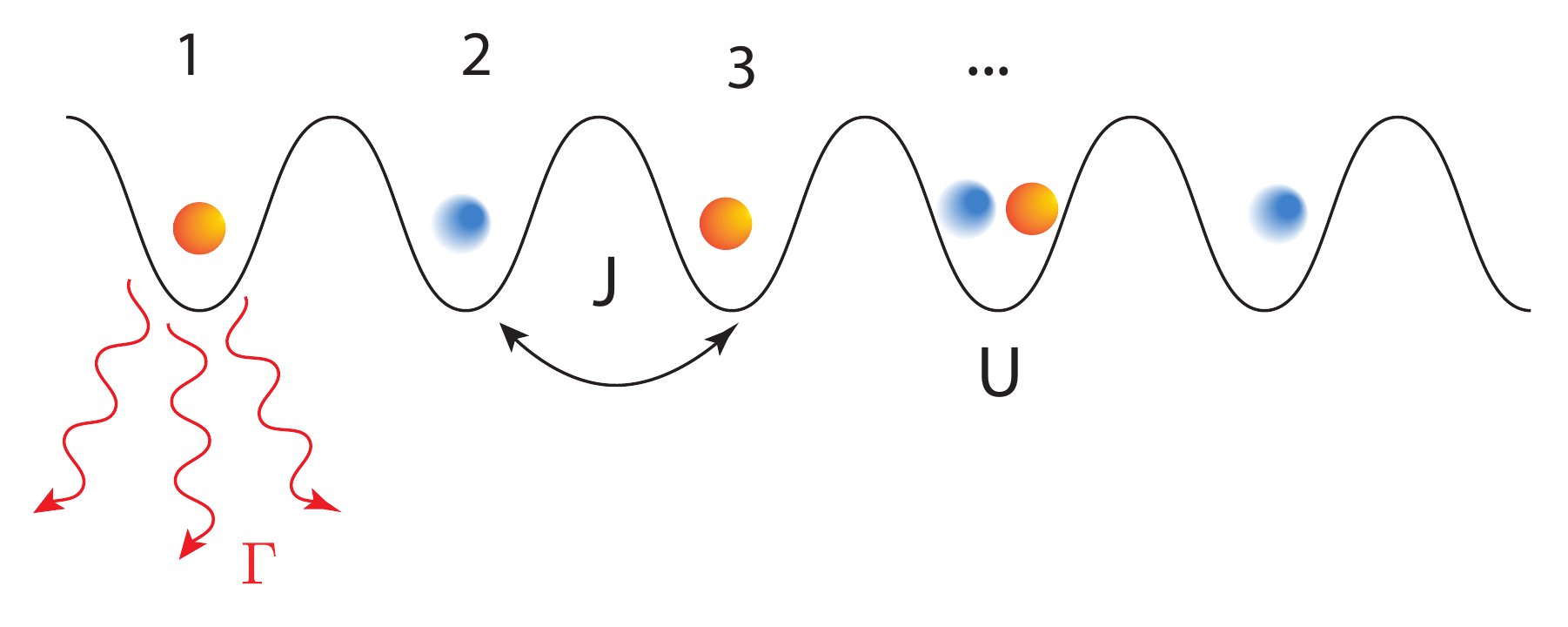}
 \caption{Sketch of the one dimensional lattice. The two spin components of the electrons  are represented by different colours. The 
 hopping between nearest-neighbour sites is labeled by $J$, while the Coulomb on-site energy is represented by $U$.
 The first site in the chain is coupled to the external reservoir and has a dissipation rate $\Gamma$.  }
 \label{fig:lattice}
\end{center}
 \end{figure}

We demonstrate the efficiency  of our approach on the $SU(2)$ fermionic  Hubbard model 
with localized  particle loss at one end of the chain, and analyze the dynamics of various observables such as 
occupations and currents.  This system, sketched in Fig.~\ref{fig:lattice}, can be relatively easily realized 
with ultracold atoms, by trapping fermions in an optical lattice, and using for example an 
electron beam~\cite{Jaksch.2008, Barontini.2013} to remove particles 
at one site. The  Hubbard model  is one of the simplest models
that describes strongly interacting particles on a lattice. It is defined by the Hamiltonian,
\begin{equation}
 H = - { J\over 2} \sum_{\sigma}\sum_{x  =1}^{N-1}\big ( c^\dagger_{x\sigma}c_{x+1\sigma} +h.c.\big)
+ U \sum_{x}  n_{x\uparrow}n_{x\downarrow}\; ,  
\label{eq:Hubbard}
\end{equation}
where $c^\dagger_{x\sigma}$ creates a fermion at site $x$ with spin $\sigma$, 
$J$ denotes the hopping amplitude between nearest-neighboring sites, $U$ represents 
the interaction energy, and $n_{x \sigma} =c^\dagger_{x\sigma}c_{x\sigma} $ stands for the number operator 
at a given site. 
In the absence of dissipation or losses, the model belongs to the class of integrable models~\cite{LiebWu,Ogata.1990,EsslerBook}. 
To induce dissipation, we couple the first site  to an external reservoir at time $t=0$, and induce  particle loss there
by the dissipators $F_{1\sigma} =  c_{1\sigma}$. SU(2) symmetry then requires that the strength of the dissipators 
$F_{1\uparrow}$ and $F_{1\downarrow}$ be equal, $\lambda = \sqrt{\Gamma}$.

We perform TEBD simulations for this model, starting from an infinite temperature state,
while benchmarking our numerical computations with exact third quantization results for the non-interacting 
case, $U=0$. The dissipator generates particle loss and a depletion region around itself, 
thereby inducing a depletion and current front, penetrating into the  Hubbard chain.
Surprisingly, both in the absence and in the presence of interactions, we observe  \emph{ballistic} front 
propagation. However, interactions renormalize the front velocity, and change the structure of the current  
profile and the spreading of the front dramatically. 

Our paper is organized as follows:  In Sec.~\ref{sec:MPS} we describe how the density matrix can be represented as an MPS by
using the vectorization procedure. In Sec.~\ref{sec:symmetries} we extend the concept of non-abelian symmetries to the Liouvillian evolution and 
construct the general non-abelian NA-MPS representation for the density matrix.   In Sec.~\ref{sec:superfermion} we introduce the superfermion 
representation, which represents a practical and useful formalism that allows to construct a new set of creation and annihilation operators for the dual space. We use this formalism to construct the Liouvillian operator for the Hubbard chain with losses.  In Sec.~\ref{sec:results} we demonstrate our approach on  a semi-infinite SU(2) Hubbard chain with losses at one end, and present numerical results for various quantities of interest such as the average 
current or average density. In the non-interacting limit we compare these averages with the third quantization results.  
In Sec.~\ref{sec:conclusions} we sumarize the results and present the conclusions of our work. For completeness, we give in Appendix~\ref{app:3rdQT} details on the third quantization treatment of the free fermion case.


\section{MPS representation of the density matrix}\label{sec:MPS_general}\label{sec:MPS}

TEBD is one of the leading approaches to simulate
time dependent correlated systems subject to local interactions.\cite{Vidal.2003, Vidal.2004} It is best suited to one-dimensional 
lattice models with a tensor product Hilbert space structure,
\begin{equation}
\mathfrak{H}   = \mathfrak{H}_1 \otimes \mathfrak{H}_2 \otimes \dots \otimes \mathfrak{H}_N \;,
 \label{eq:H}
\end{equation}
where $\mathfrak{H}_j$ is the Hilbert space associated with a single site along the chain, and  $N$  the length of the chain. 
Although it is not necessary,  we  assume for simplicity in what follows that all  sites along the chain are identical 
and possess a Hilbert space of dimension $\dim(\mathfrak{H}_j) = d$.  We are interested in simulating
the time evolution of an open system, described by the density matrix $\rho (t)$, the dynamics of which is dictated
by the Lindblad equation,\cite{Lindblad.1976, Petruccione.2002,Manzano.2020}  Eq.~\eqref{eq:Lindblad}.

The density matrix $ \rho$ as well as  other operators 
acting on $\mathfrak{H} $ are elements of an enlarged Hilbert space known as Liouville space~\cite{Petruccione.2002, Bolanos.2015}, 
$\fL=\fH\otimes \fH^*$, where $\fH^*$ represents 
the  dual ('bra') Hilbert space with respect to $\fH$.  Similar to \eqref{eq:H}, the Liouville space 
can be decomposed as a tensor product
\begin{equation}
 \fL = \fL_1 \otimes  \fL_2 \otimes \dots \otimes  \fL_N \;,
 \label{eq:L}
\end{equation}
and, by construction, the dimension  $\dim(\fL_j)=\dim(\fH_j)^2=d^{2}$. The Liouville space of operators
is endowed by a natural scalar product, $( a,  b) \equiv \textrm{Tr}_{\fH}\{ a^\dagger  b\}$.
Superoperators~\cite{Petruccione.2002} such as the Liouvillian or the dissipators are linear operators acting on $\fL$, 
and we denote them  by calligraphic letters. 

The vectorization, also termed as the Choi-Jamiolkowski isomorphism~\cite{Dzhioev.2011,Jiang.2013,Shallem.2015}, 
 is a basis dependent procedure, which  allows us to treat the Liouville space and thus the space of density operators 
as a simple vector space.  If $\{\ket{\sigma_x}\}$  is a local basis spanning the Hilbert space $\fH_{x}$, then the operators 
$\ketL{\sigma_x\tsigma_x} \equiv \ket{\sigma_x}\,\bra{\tsigma_x} $ form a basis in 
$\fL_x$,~\footnote{Notice that, to differentiate between a state that belongs to $\cH$ or $\cL$ we use slightly different ket-bra notations. }
 and we can expand the density operator in this basis.  
 
The Lindblad equation~\eqref{eq:Lindblad} is then transformed 
 into a regular Schr\" odinger-like equation, 
\begin{equation}
i\frac{d}{dt}\ketL\rho = \mathbb{L} \ketL \rho \, , 
\label{eq:evolution}
\end{equation}
with   $\it{\mathbb{L}}$ denoting  the vectorized superoperator, 
\bea
\mathbb{L} &=& H\otimes I -I\otimes H^* 
\label{eq:construction}
\\
&+& i \, \textstyle{\sum_{F} } \, \lambda_F \,\big\{ 2 F\otimes F^* - F^\dagger F\otimes I-I\otimes (F^\dagger  F)^*\big\},
\nonumber
\eea
with $I$ the unit operator over $\fH$.  Eq.~\eqref{eq:evolution} is then formally  integrated as ~\footnote{Within the time evolution we set $\hbar=1$.} 
\begin{equation}
\ketL{\rho (t)} =e^{-i \,\mathbb{L} \,  t}  \ketL{\rho(0)} \, . 
\label{eq:rho_t}
\end{equation}

%
%
%

As long as the Hamiltonian $H$ and the dissipators $F$ are local,  we can use the efficient 
methodology of matrix product states to generate the time evolution. In fact, similar
to matrix product states, we can rewrite the state $\ketL{\rho}$  in an MPS form, 
\begin{eqnarray}
 \ketL{\rho} &=& \sum_{a_{1}, \dots a_{N-1}} \sum_{\sigma_1, \dots \tsigma_L} 
 \mathcal{R}^{[1] \, a_1}_{(\sigma_1\tsigma_1)} \mathcal{R}^{[2] \, a_2}_{a_1 (\sigma_2\tsigma_2)} \dots \mathcal{R}^{[N]}_{a_{N-1} (\sigma_N\tsigma_N)} \times\nonumber \\
 & & \ketL{\sigma_1\tsigma_1} \otimes \ketL{\sigma_2\tsigma_2} \otimes \dots \otimes \ketL{\sigma_N\tsigma_N} \; .
 \label{eq:left_MPS_nonsym_full}
\end{eqnarray} 
%
Notice that the evolution operator $\mathbb{V}(t) =e^{-i \,\mathbb{L}\, t}$ is non-Hermitian~\cite{Ashida.2020} due to dissipation introduced by the jump operators. 
Nevertheless, once $ \ketL{\rho} $ is rewritten in this matrix product form, we can use the MPS machinery to generate 
the time evolution  of $\ketL{\rho(t)}$ within the TEBD approach,\cite{Zwolak.2004, Verstraete.2004} just as for
unitary wave function evolution. As we mentioned in the introduction, the only problem arises due to dimensionality, since 
for just two fermionic degrees of freedom, the dimension of the local Liouville space $\fL$
is already 16. In the following, we shall reduce this number using non-Abelian  symmetries down to 10, a number that 
can already be handled with standard desktop computers or work stations. 

To summarize the notations, we shall use capital latin letters $A$, to label operators acting over the Hilbert space $\fH$. 
States in $\fH$ are denoted by the standard Dirac ket notation $\ket{s}$, with their  Hermitian conjugates referred to as
 $\bra{s}$. The complex conjugation of an operator $A$ with respect to computational basis is denoted as $A^*$.
  With calligraphic letters, e.g. $\cal{A}$,
we denote superoperators acting over the Liouville space, $\fL$. 
Double-line letters, $\mathbb{A}$, denote operators acting on the vectorized Liouville space, while  round kets $\ketL{\rho}$ denote
 states in the vectorized  Lindblad space.

\subsection{Symmetries in the  Liouvillian approach}\label{sec:symmetries}
In this section, we extend the non-abelian TEBD (NA-TEBD) approach of Ref.~\onlinecite{Werner.2020}
 to  Lindbladian evolution and show that – with certain extensions –
most concepts of Ref.~\onlinecite{Werner.2020} remain applicable. 
In Hamiltonian systems, the symmetries are represented by 
a set of unitary (or antiunitary) transformations, $U(g)$, which   leave the  Hamiltonian
invariant for all elements $g$ of the symmetry group $G$,
\be
U(g)\,H\,U^\dagger(g) = H\,, \phantom{n}\textrm{ for} \phantom{n}\forall\,  g\in G\;,
\label{eq:invariance}
\ee 
or, equivalently $ U(g) H= H\, U(g)$. 
These symmetry operations are  naturally extended to the Liouville space $\fL$ by the superoperator, 
\be
 {\cal U} (g)  \;:  \phantom{nn} \rho  \phantom{n} \to  \phantom{n}  U(g)  \,  \rho \,  U^\dagger (g)\;. 
\ee
The superoperators ${\cal U}(g) $ thus represent a \emph{symmetry} of the Liouvillian  if 
\be
{\cal U}(g) \,{\cal L} =  {\cal L}\,\,  {\cal U}(g) \;,
\label{eq:Liouville_symmetry}
\ee 
written equivalently  as Eq.~\eqref{eq:Lsymmetry}, when applying both sides to an arbitrary density operator. 
Correspondingly, a generator $ J$ of some continuous symmetry, $U=e^{i \phi J}$, is represented  in 
Liouville space by a superoperator $\cal J$ as 
\be
 J \phantom{n}\to \phantom{n}{\cal J}[\, \rho\,] \equiv  J \,  \rho -  \rho \,  J^\dagger\;,
\ee
where we now allowed complex parameters, $\phi$, and corresponding 
non-Hermitian generators. 

Clearly, for Hamiltonian dynamics, ${\cal L} \to  {\cal L}_H = [H\,,\,.\,]$, 
the condition \eqref{eq:Liouville_symmetry}, 
$[{\cal L}_H,{\cal U}] = 0$ and the familiar symmetry condition  $ [H\,,\, U] = 0$ are equivalent. For open systems, however, 
$[{\cal L},{\cal U}] = 0$ implies additional constraints on the structure of dissipators. 
Dissipators  are operators, and as such,  formally elements of the Liouville space, $F\in \fL$. 
Similar to states in the Hilbert space, the Liouville space of operators can  be devided 
into irreducible subspaces of groups of irreducible tensor operators, $F_q$, which transform according 
to some irreducible representation of the symmetry group, 
\be 
{\cal U}(g)\,F_{q} =  {U(g)}\,F_{q} \,U(g)^\dagger = \sum_{q^\prime }\Gamma(g)^{q^\prime}_q\, F_{q^\prime}\;,
\ee
with $\Gamma(g)$ the  representation of the group element, $g$. It is easy to show that the symmetry operation 
${\cal U}$ commutes with the action of the dissipator under the condition that 
the strength of the dissipators belonging to the same irreducible representation is equal, 
\be
{\cal D}\sim \lambda \sum_q {\cal D}_{F_q}\;.
\label{eq:dissipator_condition}
\ee
This simple condition can also be naturally derived in case we construct the dissipator, as usual, 
from coupling the operators $F$ to some fluctuating fields, $\varphi_F$, and request that 
the subsystem and its environment be invariant under symmetry operations, as a composite system.

In the following, we shall assume that our dissipators satisfy Eq.~\eqref{eq:dissipator_condition}, 
and focus furthermore on local symmetry operations, when $U(g)$ factorizes as
\begin{equation}
U(g)=U_1(g)\otimes U_2(g)\otimes\dots \otimes U_{N}(g)\;.
\end{equation}
In this case, the   Hilbert space  can be decomposed 
into multiplets at each site
\be
 \fH_j = \mathrm{span}\left\lbrace \ket{\Gamma^{\rm local}; \tau,\mu} \right\rbrace_j \;,
\label{eq:Hilbert_space_span}
\ee
with index $\Gamma^{\rm local}$ denoting the irreducible representation,  $\tau$ 
running over multiplets with a given representation, and $\mu$ labeling internal states of a multiplet.  
In general, if the model displays $n_S$ commuting symmetries,
the  label $\Gamma$ becomes a vector $\Gamma=(\Gamma_1, \dots, \Gamma_{n_S})$, where 
$n_S$ is the total number of commuting symmetries. For the Hubbard model, studied here, e.g., we  use 
spin SU(2) and charge U(1) symmetries, corresponding to $n_S=2$, and  $\Gamma$  then refers to the 
corresponding spin and charge quantum numbers, $\Gamma\to (S,N)$. Having classified local states by 
symmetries, states in the Hilbert space $\fH$ can then be represented as non-Abelian 
matrix product states\cite{Werner.2020, Weichselbaum.2020}
where the matrix product representation is decomposed into a trivial Clebsch-Gordan layer, and a non-trivial layer
of reduced dimension, containing all necessary information. 

This concept can be carried over to the Liouville space, while here we shall do that by implementing non-Abelian symmetries 
in vectorized space of operators. In the vectorized space, where the symmetry 
superoperator $\cal U$ is represented   as ${\cal U}\to \mathbb{U} =  U \otimes  U^*$, 
while the symmetry generators become
\be 
\mathbb{J} = J\otimes I-I\otimes J^*\, .
\label{eq:supergenerators}
\end{equation}
Irreducible tensor operators $T_q$, as well as the dissipators $F_q$, are represented as states $T_q\to |T_q)$, which then obey 
\be
 \mathbb{U}(g)|T_q) =   \sum_{q^\prime }\Gamma(g)^{q^\prime}_q\, |T_{q^\prime})\;.
\ee
Similar to Eq.~\eqref{eq:Hilbert_space_span}, the local vectorized Liouville space  can be organized into  'operator multiplets', 
$T^{[j]}_q$, represented in the local vectorized space as $ \ketL {\bbGamma^{\rm loc}; T , q}_j $, where we use the symbol 
$\bbGamma$ to emphasize that these states belong to the vectorized Liouville space.

\begin{figure}[t!]
\begin{center}
 \includegraphics[width=0.95\columnwidth]{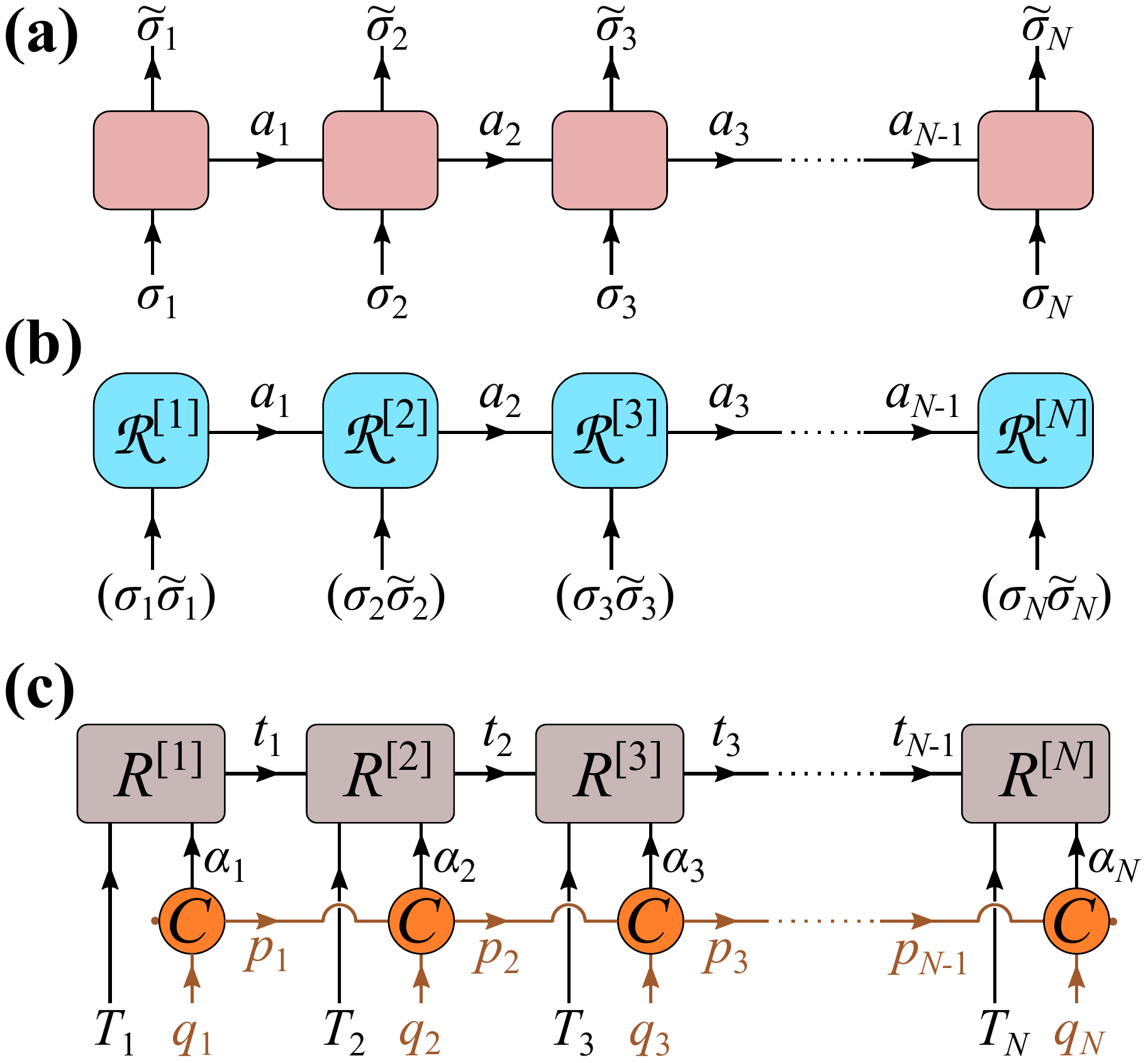}
 \caption{{(a)  MPO representation of a general density matrix.  (b) Merging the legs of the MPO allows us 
 to  represent the density matrix as an MPS.  
 (c) Graphical representation of the NA-MPS in equation~\eqref{eq:NAMPS}. The NA-MPS tensor consists of two layers. 
 The upper layer is the 'core'  NA-MPS, while the lower layer is built from Clebsch-Gordan coefficients
 and  carries all symmetry-related information. }
 }\label{fig:NAMPS}
\end{center}
 \end{figure}

We can now extend the  NA-TEBD approach to the vectorizes space and  represent the density operator 
as a non-abelian MPS (NA-MPS)~\cite{Werner.2020}
\begin{widetext}
\begin{eqnarray}\label{eq:NAMPS}
\ketL{\rho} = \sum_{\brc{\bbGamma^{\loc}_l}} \sum_{\brc{\bbGamma_l}} \sum_{\brc{ t_l}} \sum_{\brc{T_l}} \; \sum_{\brc{\alpha_l}} & & R^{[1]}(\brc{\bbGamma}^{[1]})_{T_1 \, \alpha_1}^{t_1} \,  R^{[2]}(\brc{\bbGamma}^{[2]})_{t_1 \, T_2 \, \alpha_2}^{t_2} \, \dots \,  R^{[N]}(\brc{\bbGamma}^{[N]})_{t_{N-1} \, T_N \, \alpha_N} \nonumber \\ 
 \sum_{\brc{ T_l}} \sum_{\brc{Q_l}} & & C(\brc{\bbGamma}^{[1]})_{0 \, q_1}^{p_1,\,\alpha_1} \, 
C(\brc{\bbGamma}^{[2]})_{p_1 \, q_2}^{p_2,\,\alpha_2} \, \dots \, C(\brc{\bbGamma}^{[N]})_{p_{N-1} \, q_N}^{0 ,\, \alpha_N} \nonumber \\ & &  
\ketL{\bbGamma^\loc_1; T_1,q_1} \otimes \ketL{\bbGamma^\loc_2; T_2,q_2} \otimes  \dots \otimes \ketL{\bbGamma^\loc_N; T_N,q_N} \; . 
\end{eqnarray}
\end{widetext}
 In our construction, the NA-MPS structure has two layers (see Fig.~\ref{fig:NAMPS}). The upper layer contains the 
 tensors $R^{[x]}(\brc{\bbGamma}^{[x]})^{t_x}_{t_{x-1} \, T_x \, \alpha_x}$, which  carry the essential information  of the state, 
 while the lower layer includes exclusively Clebsch-Gordon coefficients, $C(\brc{\bbGamma}^{[x]})_{p_{x-1} \, q_x}^{p_x \, \alpha_x}$,
  and carries all the symmetry-related 'trivial' information.\cite{Werner.2020}
%
%
%
%
%
%
%
 In Eq.~\eqref{eq:NAMPS} we used a compact notation for the set of   irreducible  representation labels, 
 $  \brc {\bbGamma}^{[x]} =  (\bbGamma_{x-1}, \bbGamma^{\loc}_x, \bbGamma_x) $, associated with the legs of the tensors $R^{[x]}$.
 The index $\alpha_l$ represents the  outer-multiplicity label, and depends in general on $\brc {\bbGamma}^{[x]}$. 
 This NA-MPS structure naturally incorporates abelian symmetries as well. 
Then Clebsch-Gordan coefficients are just 1 for representation labels ${\bbGamma}^{[x]}$ allowed by the selection rules.

In case of local  dissipators and a local Hamiltonian containing only nearest neighbor 
 interactions and hopping, we can now proceed just as in Ref.~\onlinecite{Werner.2020}, and 'Trotterize' 
 the (vectorized) evolution operator, $\mathbb{V} = e^{-i\,t\, \mathbb{L}}$, and eliminate the 
 Clebsch-Gordan layer.\cite{Werner.2020} TEBD or DMRG steps can then be carried out very efficiently, 
by carrying out the singular value decompositions in separate symmetry sectors independently.

\section{Vectorization: superfermion representation }\label{sec:superfermion}

We now vectorize the Liouvillan space using the  so-called superfermion representation,\cite{Dzhioev.2011,Dzhioev.2011a, Dzhioev.2012}
which introduces a new set of creation/annihilation operators $ \tc_{j\sigma}$ acting in the dual Fock space.
These operators satisfy the usual anticommutation relations,
$ \{\tc_{j\sigma},\tc^\dagger_{k\sigma'}\}=\delta_{jk}\delta_{\sigma\sigma'}$,  etc. 
The operators  $c_{j\sigma}^{(\dagger)}$ and $\tc^{(\dagger)}_{k\sigma'}$ act nontrivially on different Fock spaces, and also anticommute, 
$ \{c_{j\sigma},\tc^{(\dagger)}_{k\sigma'}\}=0$.
In this formalism the vectorized Liouvillian \eqref{eq:construction} becomes
\begin{eqnarray}
\mathbb{L}  &=& -{J\over 2}\sum_{\sigma}\sum_{i=1}^{L-1}\big (  c^\dagger_{i\sigma}c_{i+1\sigma} +h.c.\big)
+ U \sum_{i}  n_{i\uparrow}n_{i\downarrow}\;\nonumber\\
& +&{J\over 2} \sum_{\sigma}\sum_{i=1}^{L-1}\big (  \tc^\dagger_{i\sigma}\tc_{i+1\sigma} +h.c.\big)
- U\sum_{i}  \tn_{i\uparrow}\tn_{i\downarrow}\nonumber\\
& +& 2i \Gamma \sum_{\sigma}  c_{1\sigma}\tc_{1\sigma} 
 -i\Gamma\sum_{\sigma}\big( \tc^\dagger_{1\sigma}\tc_{1\sigma}-c^\dagger_{1\sigma}c_{1\sigma} \big)\, .
   \label{eq:L_superoperator}
\end{eqnarray}

Here we follow  a general approach, and  form the local tensor product space by acting 
with the $c_i^{(\dagger)}$ and  $\tilde c^{(\dagger)}_i$  operators.  
In the specific case studied here, however, we  may consider  regarding the Liouvillian 
Eq.~\eqref{eq:L_superoperator}  as a non-Hermitian Hamiltonian for a two-chain (ladder) 
model coupled via the first site, and construct an MPS representation by separating the sites and 
ordering the  dual operators $\tilde c^{(\dagger)}_i$ to act on the sites 
$i=-N+1,-N+2,\dots, 0$, while the regular operators  $c_i^{(\dagger)}$  act on the sites $i=1, 2\dots N-1, N$ 
along the chain. In this way,  we would double the length of the chain and the tilde and the regular sites
 are coupled at the center of the chain. The price we would pay, however, is that even an infinite temperature initial state
 would contain long-ranged entanglement. Moreover, in the general case of more local  dissipators along the chain,  
 dissipators would induce long-ranged terms in the effective Hamiltonian, causing additional difficulties.
In contrast, within the tensor product state approach we follow, the size of the local space is larger, but 
the initial state has a simple structure, dissipators remain local, and the chain is half as long as in the 
unfolded chain approach. 

The vectorized Liouvillan, Eq.~\eqref{eq:L_superoperator} then  acts 
on the Fock space of vectorized operators, i.e., the vectorized Liouville space. 
To propagate the vectorized Linblad equation, Eq.~\eqref{eq:rho_t}, we thus need to represent  
the  density matrix  $\ketL {\rho (t)}$  within the superfermions' Fock space, and then  follow the regular TEBD approach,
generated by Eq.~\eqref{eq:L_superoperator}.
 
Let us now elaborate on  the symmetries of the Hamiltonian~Eq.~\eqref{eq:Hubbard} and their representation on the 
Liouville space.  The  spin operator ${\bf S} = {1\over 2} \sum_{i,\sigma\sigma'} c^\dagger_{i\sigma}\boldsymbol {\sigma}_{\sigma
\sigma'} c_{i\sigma'}$ as well as the normal ordered total charge $N = \sum_{i,\sigma}(c^\dagger_{i\sigma}c_{i\sigma}-1)$ 
commute with $ H$ in~\eqref{eq:Hubbard}, which  thus  displays a $ G  = SU (2) \otimes U (1) $ symmetry. Accordingly, 
local states in the Hilbert space $\fH^\loc$ can be classified  by spin and charge  quantum numbers,
  $\Gamma^\loc = (S ^\loc, N^\loc)$.

To carry out a similar  classification in the local, vectorized Liouville space, 
we must represent first symmetry operations acting on this space, as outlined earlier. 
Spin rotations, e.g., are generated by the operators
$$
\mathbb{S}_j^\alpha = S_j^\alpha \otimes I-I\otimes {S_j^\alpha}^*\, ,\phantom{aaa} (\alpha = x,y,z)\;.
 $$
 Similarly, 
 $U(1)$ gauge transformations are generated by $\mathbb{N}_j = N_j \otimes I-I\otimes {N_j}$ in the vectorized space.
 Notice that the 'raising operator' $\mathbb{S}^+$ becomes 
 $\mathbb{S}^+ =\mathbb{S}^x  + i \,\mathbb{S}^y = S^+ \otimes I-I\otimes {S^-}$ when acting over the vectorized space.
 
 We can easily represent the operators  $\mathbb{S}^\pm$ and $\mathbb{S}^z$  in  the superfermion representation
by using the operators $c_{j\sigma}$ and $\tilde c_{j\sigma}$ as\cite{Dzhioev.2011,Dzhioev.2011a}
\begin{eqnarray}\
{\mathbb{S}^+_j } &= &  c^\dagger_{j\uparrow}c_{j\downarrow} - \tc^\dagger_{j\downarrow}\tc_{j\uparrow}\;, \label{eq:spin_generators}\\
{\mathbb{S}}^z_j &= & {1\over 2}\big (c^\dagger_{j\uparrow}c_{j\uparrow}-c^\dagger_{j\downarrow}c_{j\downarrow}\big )-
 {1\over 2}\big (\tc^\dagger_{j\uparrow}\tc_{j\uparrow}- \tc^\dagger_{j\downarrow}\tc_{j\downarrow} \big )\;,\nonumber\\
{\mathbb{S}^-_j } &= &c^\dagger_{j\uparrow}c_{j\downarrow} - \tc^\dagger_{j\downarrow}\tc_{j\uparrow} \; . \nonumber
\end{eqnarray}
In a similar way, the U(1) symmetry related to  particle conservation is  generated by 
\begin{gather}
\mathbb{N}_j = \sum_{\sigma} (c^\dagger_{j\sigma}c_{j\sigma}-\tc^\dagger_{j\sigma}\tc_{j\sigma})\,.
\label{eq:charge_generators}
\end{gather}

Having identified the symmetry generators over the vectorized Liouville space, we can now identify 
families irreducible tensor operators. 
In the local  Liouville space, $\fL_j$ at site $j$, e.g., 
the spin operators $-S_j^+/\sqrt{2}$, $S^z$, $S_j^-/\sqrt{2}$, form the three components
of a spin ${\cal S}=1$ vector operator of charge ${\cal N}=0$. Similarly, the 
 creation  operators are $c^\dagger_j = (c^\dagger_{j\uparrow},c^\dagger_{j\downarrow})$
 and the annihilation operators $\tc_j = (\tc_{j\uparrow},-\tc_{j\downarrow} )$
 are spin $ {\cal S}=1/2$ operators of charge ${\cal N}=1$  and ${\cal N}=-1$, respectively.  

\begin{table}[b!]
\begin{tabular}{|c| c| c| c|}
 \hline
 $\bbGamma =(\mathbb{S},\; \mathbb{N}) $ &  $\dim(\bbGamma)$ & $ T$ & states  \\
 \hline \hline
 (0, -2) & 1 & 1 & $\tc^\dagger_{\uparrow}\tc^\dagger_{\downarrow} \ketL{0}$ \\ 
 & & & \\	
 (0, 0) & 1 & 1 & $\ketL{0}$ \\		 
 & & 2& ${1\over \sqrt{2}} (c^\dagger_{\uparrow}\tc^\dagger_{\uparrow}-c^\dagger_{\downarrow} \tc^\dagger_{\downarrow} )\ketL{0}   $ \\
 & & 3& $c^\dagger_{\uparrow} c^\dagger_{\downarrow}\tc^\dagger_{\uparrow}\tc^\dagger_{\downarrow} \ketL{0}$\\
 & & & \\
 (0,\; 2) & 1 & 1 &  $c^\dag_{\uparrow} c^\dag_{\downarrow} \ketL{0}$ \\ 
 & & &\\
 (${1\over 2}$, -1) & 2 & 1 & $\big\{\tc^\dagger_{\uparrow}\ketL{0}$, $\tc^\dagger_{\downarrow} \ketL{0}\big\}$ \\ 
 &  &  2 &  $\big\{c^\dagger_{\uparrow}\tc^\dagger_{\uparrow}\tc^\dagger_{\downarrow}\ketL{0}$, 
 $c^\dagger_{\downarrow}\tc^\dagger_{\uparrow}\tc^\dagger_{\downarrow} \ketL{0}\big\}$\\
 & & & \\
 (${1\over 2}$ ; 1) & 2 & 1 & $\big\{c^\dag_{\uparrow} \ketL{0}$, $c^\dag_{\downarrow}\ketL{0}\big\}$\\ 
 &  & 2  & $\big\{c^\dag_{\uparrow} c^\dag_{\downarrow}\tc^\dagger_{\downarrow} \ketL{0}$, $c^\dag_{\uparrow} c^\dag_{\downarrow}\tc^\dag_{\uparrow}\ketL{0}\big\}$\\ 
 & & & \\
 & & & $\big\{\;c^\dagger_{\downarrow}\tc^\dagger_{\uparrow}\ketL{0} $\\ 
(1, 0)&3 & 1& ${1\over \sqrt{2}} (c^\dagger_{\uparrow}\tc^\dagger_{\uparrow}+c^\dagger_{\downarrow} \tc^\dagger_{\downarrow} )\ketL{0}   $ \\
 & & & $c^\dagger_{\uparrow}\tc^\dagger_{\downarrow}\ketL{0}\;\big\} $\\ 
 & & & \\ 
  \hline
\end{tabular}
\caption{$SU(2)$ multiplets as they are indexed by the quantum numbers for the total spin $ \mathbb{S} $ and  occupation $ \mathbb{N} $.}\label{tab:fermion_site}
\end{table}

In the vectorized form, all these 
operators of the Liouville space $\fL_j$ are represented  as  Fock states. The spin 
operator $S^+_j \in \fL_j$ is identified, e.g.,  by the state $|S^+_j ) = c^\dagger_{j\uparrow} \tilde c^\dagger_{j\downarrow} |0) $.
The $4\times 4 =16$  local operators are then represented by  16  Fock states, listed 
in Table~\ref{tab:fermion_site},  which can then be organized into 6 symmetry sectors, and altogether 10  multiplets. 

Notice that, although the Lindbladian dissipator removes particles, the Lindbladian  $U(1)$ charge ${\cal N}$
(or  $\mathbb{N}$ in the vectorized form) is conserved. This can be verified directly by 
investigating the condition \eqref{eq:Lsymmetry}, or by looking at the commutator of 
Eqs.~\eqref{eq:charge_generators} and \eqref{eq:L_superoperator}. We can thus use 
the fulll $SU(2)\times U(1)$ weak  Liouvillian symmetry, even though the total charge is not conserved.

\section{Application to the SU(2) Hubbard model }\label{sec:results}
%
%

In this section, we demonstrate our approach on the semi-infinite fermionic SU(2) Hubbard model with a particle sink 
at the end of the chain. We  execute a 'dissipation quench': we start our simulations by  constructing 
a half-filled infinite temperature state, which  can be created relatively simply in cold atom experiments,\cite{Braun.2013, Tarruell.2018} 
and turn on a 'particle sink' process at site $0$  at time $t=0$. The sink empties the semi-infinite chain, thereby creating 
a moving front between the emptied and occupied regions, and the quantum-mechaical time evolution 
generating entanglement. 
 
In the vectorized formalism, the  infinite temperature half-filled initial state translates to the state
\begin{gather}
\ketL{\rho(0)} =\prod_{x=1}^{N} \frac {1+c^\dagger_{x\uparrow}c^\dagger_{x\downarrow}\tilde c^\dagger_{x\uparrow}\tilde c^\dagger_{x\downarrow} 
+ c^\dagger_{x\uparrow}\tilde c^\dagger_{x\uparrow} + c^\dagger_{x\downarrow}\tilde c^\dagger_{x\downarrow} } 2 \;\ketL{0}\, .
\label{eq:rho_0}
\end{gather}
Being the product of local states, this vectorized state  has   a trivial  MPS representation,  
 and possesses Liouvillian quantum numbers ${\cal S} = 0$, and ${\cal N} =0$,   implying 
  that $\ketL{\rho(0)}$ is an SU(2) super-singlet.  

To compute time dependent observables, we simply need to evolve the density operator ${\rho(0)} $ in time. 
The time evolution of the expectation value of an operator $A$ is then calculated 
from the time dependent density matrix $\rho(t)$ as the left vaccuum vector for the Liouvillian operator 
\begin{equation}
\label{eq:A_t}
\average{A(t)}= \mathrm{Tr}_{\rm S}\brc{A\rho(t)}\;. 
\end{equation}
The normalization of the density matrix implies $\mathrm{Tr}_{\rm S}\brc{\rho(t)}=1$. Within the vectorized formalism,
this translates into  $ ( I \ketL{\rho(t)} =1$, where $\ketL{I}$ can be associated, up to  normalization and a phase, to the density matrix 
of the infinite temperature state. The latter lacks any coherence and has only equal, diagonal elements. 
Within the superfermion representation
this state can be explicitly constructed as~\cite{Dzhioev.2011}
\begin{equation}
\ketL{I} = \exp \big \{ -i \sum_{x,\sigma}c^\dagger_{x, \sigma}\tilde c^\dagger_{x, \sigma}  \big\} \ketL{0}.
\end{equation} 
The average~\eqref{eq:A_t}  can then be cast into the matrix element 
\begin{equation}
\average{A(t)} = \braL{I} \mathbb{A}\ketL{\rho(t)}\;.
\end{equation}
We computed $\ketL{\rho(t)}$ using the non-Abelian time-evolving block decimation (NA-TEBD) method, following the lines of  
 Ref.~\onlinecite{Werner.2020}, applied there for Hamiltonian time evolution. The crucial difference here is that now time evolution is non-Hermitian, 
and is performed in the vectorized space.  To benchmark NA-TEBD, 
 we also determined  the time evolution of the non-interacting  system,  $U=0$,  by the third quantization ($3^{\rm rd}\rm QT$) approach 
 of Ref.~[\onlinecite{Prosen.2008}],  which allows us a 'numerically exact' computation of the expectation values of various operators and their
correlators in case of non-interacting  Hamiltonians.\cite{Kos.2017}
For completeness,   the $3^{\rm rd}\rm QT$ approach is briefly outlined in Appendix~\ref{app:3rdQT}.
\begin{figure}[t!]
\begin{center}
\includegraphics[width=1.0\columnwidth]{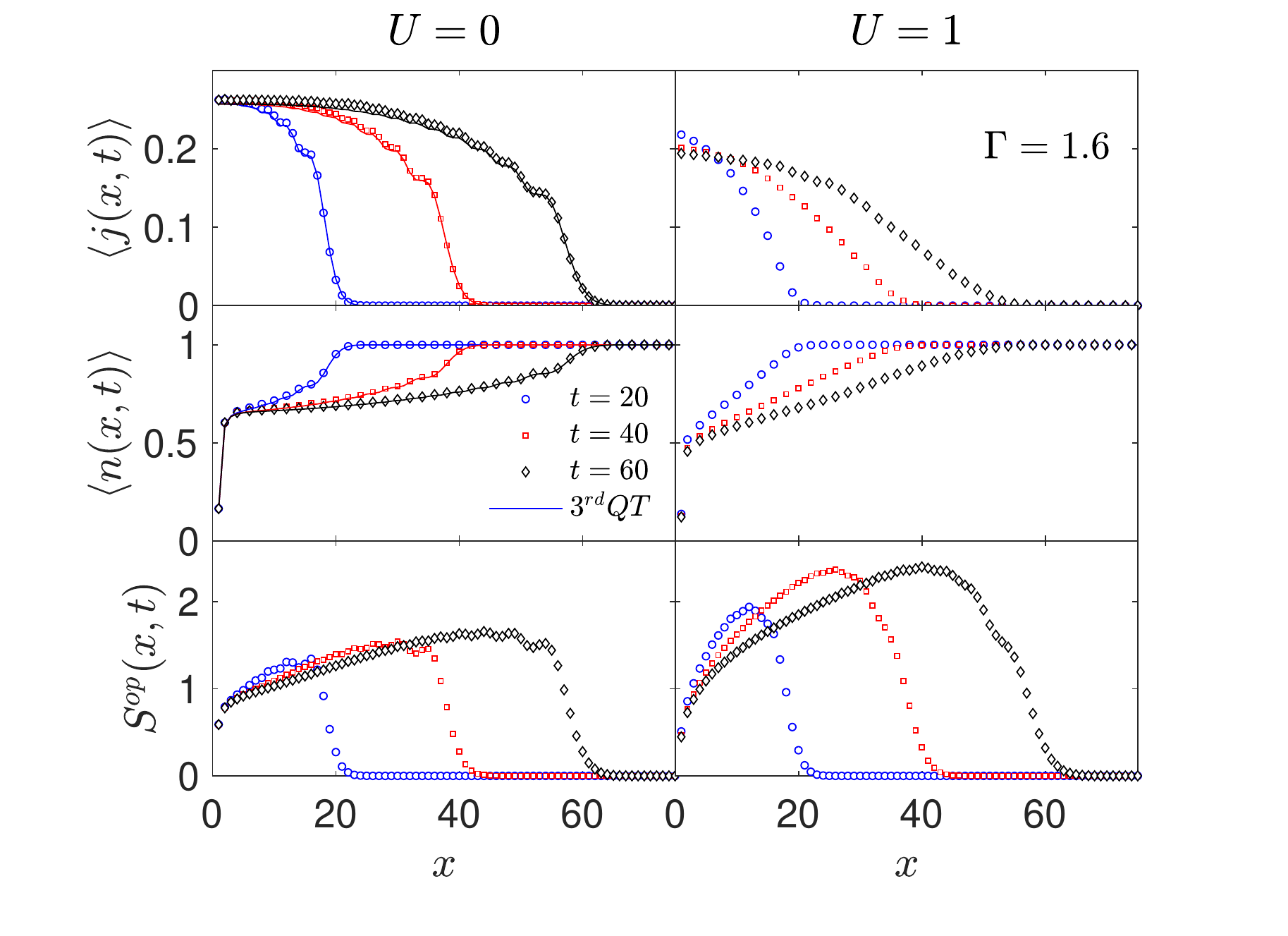}
 \caption{ Position dependence of the average current $\langle j(x,t)\rangle$, occupation $\langle n(x,t)\rangle$,
 and operator entanglement entropy $S^\text{op} (x,t)$ for various times, indicating a balistic propagation of the front,
 irrespective of the interaction strength.  
 For $U=0$, the average current $\average{j(x,t)}$
 and occupation $\average{n(x,t)}$, obtained with the $3^{rd}QT$ are displayed with solid lines.
 Bond dimension were fixed to $M=500$, and the system size is $N=100$. }
 \label{fig:curr_occup}
\end{center}
 \end{figure}

\begin{figure}[t!]
\begin{center}
 \includegraphics[width=0.95\columnwidth]{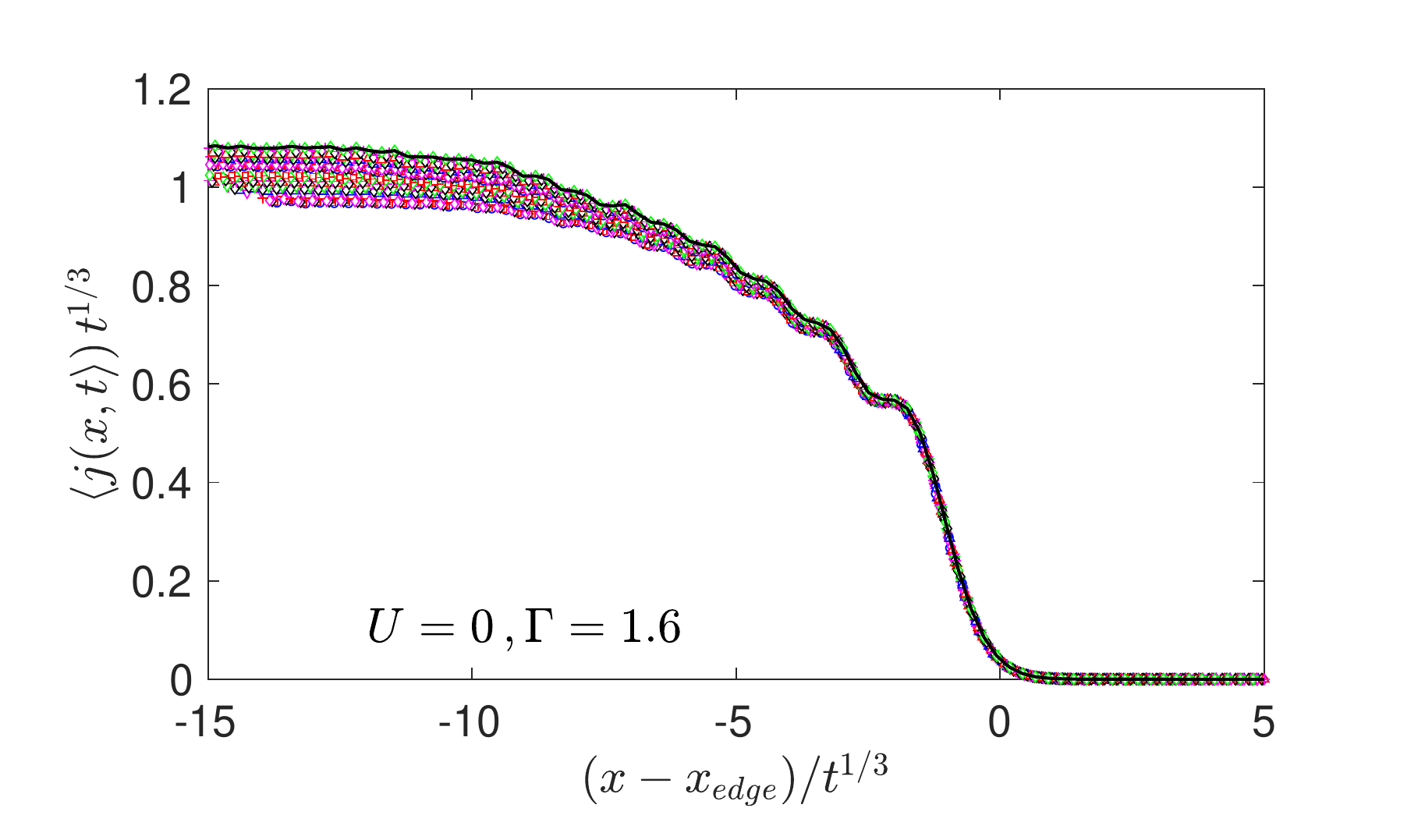}
 \caption{ Scaling collapse of the current profile near the edge  of the front, $x_\text{edge}$, for different times showing 
a staircase structure in the non-interacting limit $U=0$.   Here we display the scaling of the front for 20 different times between 
$t_1=25$ and $t_2=35$ equally spaced at an interval $\Delta t = 0.5$.}
 \label{fig:front_U=0}
\end{center}
 \end{figure}

{First we focus on the} position dependent occupation, $n(x)$, and the  current $j(x)$  between neighboring sites $x$ and $x+1$, 
defined as
\begin{align}
n(x)&\equiv\sum_{\sigma} c^\dagger_{x\sigma}c_{x\sigma}\label{eq:occup_op}\,,\\
j(x) &\equiv -i\sum_{\sigma}(c^\dagger_{x\sigma} c_{x+1\sigma}-c^\dagger_{x+1\sigma}c_{x\sigma})\label{eq:current_op}. 
\end{align}
The profiles of  $\average{j(x, t)}$ and $\average{n (x, t)}$  
are displayed  in Fig.~\ref{fig:curr_occup} at different times. The data show a qualitative difference 
between the non-interacting and interacting cases.

\begin{figure}[b!]
\begin{center}
 \includegraphics[width=0.85\columnwidth]{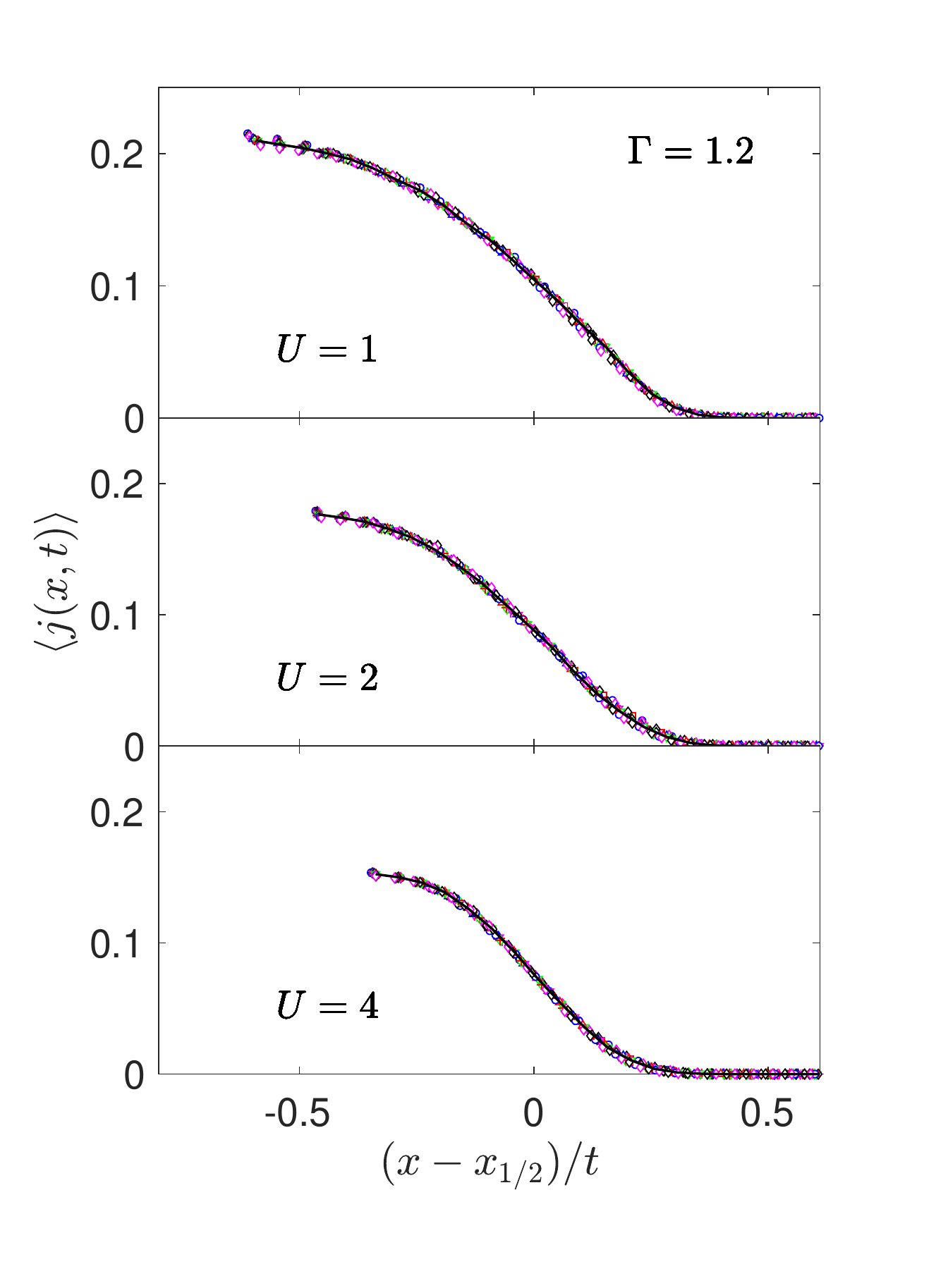}
 \caption{ Scaling collapse of the  average current profiles $\average{j(x,t)}$  indicating ballistic propagation
 and a linear in time broadening of the front 
 for any finite $U$ near the $x_{1/2}$ front edge.  In each panel we display the scaling of the front for 20 random times between 
$t_1=10$ and $t_2=35$. }
 \label{fig:front_U_ne_0}
\end{center}
 \end{figure}
 
In the non-interacting case,  $U=0$, the current profile displays a staircase-like structure around the edge of the front, 
similar to the fronts observed in free fermion systems, evolved from a state
with a density step-like initial condition~\cite{Eisler.2013}. As obvious from the first panel  on the left in Fig.~\ref{fig:curr_occup}, 
the front spreads ballistically with a velocity $v_0 = J = 1$, i.e., the maximal velocity of free quasiparticles. 
At the same time, we observe a $\sim t^{1/3}$ broadening of the front, characteristic of free Fermions.~\cite{Eisler.2013,Bertini.2016}
This is demonstrated in Fig.~\ref{fig:front_U=0}, displaying the appropriately rescaled $\average{j(x,t)}$ curves at the edge 
of the current profile, $x_{\rm edge}$, where the current starts to have a finite value. 
As the current profile evolves in time, it assumes a universal shape, and develops a fine staircase-structure, 
 indicative of free, ballistically moving particles.

For finite interactions, $U>0$,  a quite different picture emerges. The front still appears to propagate ballistically, with a somewhat 
reduced velocity, however, the current profile lacks the staircase structure, and the whole profile seems to acquire a universal shape 
 rather than just the front. This hydrodynamics-like time evolution is demonstrated in Fig.~\ref{fig:front_U_ne_0},
 where we rescale the front simply by the propagation time, $t$, around the position   $x_{1/2}$, where the average current is at half 
 maximum,  $\average{j(x_{1/2},t)} = {1\over 2} \mathrm{max}_x(\average{j(x,t)})$. 
 Notice that the current profile is a function depends on $U$;
its height as well as its width is largely suppressed  increasing $U$, thus the overall current as well as the front velocity 
are suppressed with increasing $U$.

\begin{figure}[t!]
\begin{center}
\includegraphics[width=0.95\columnwidth]{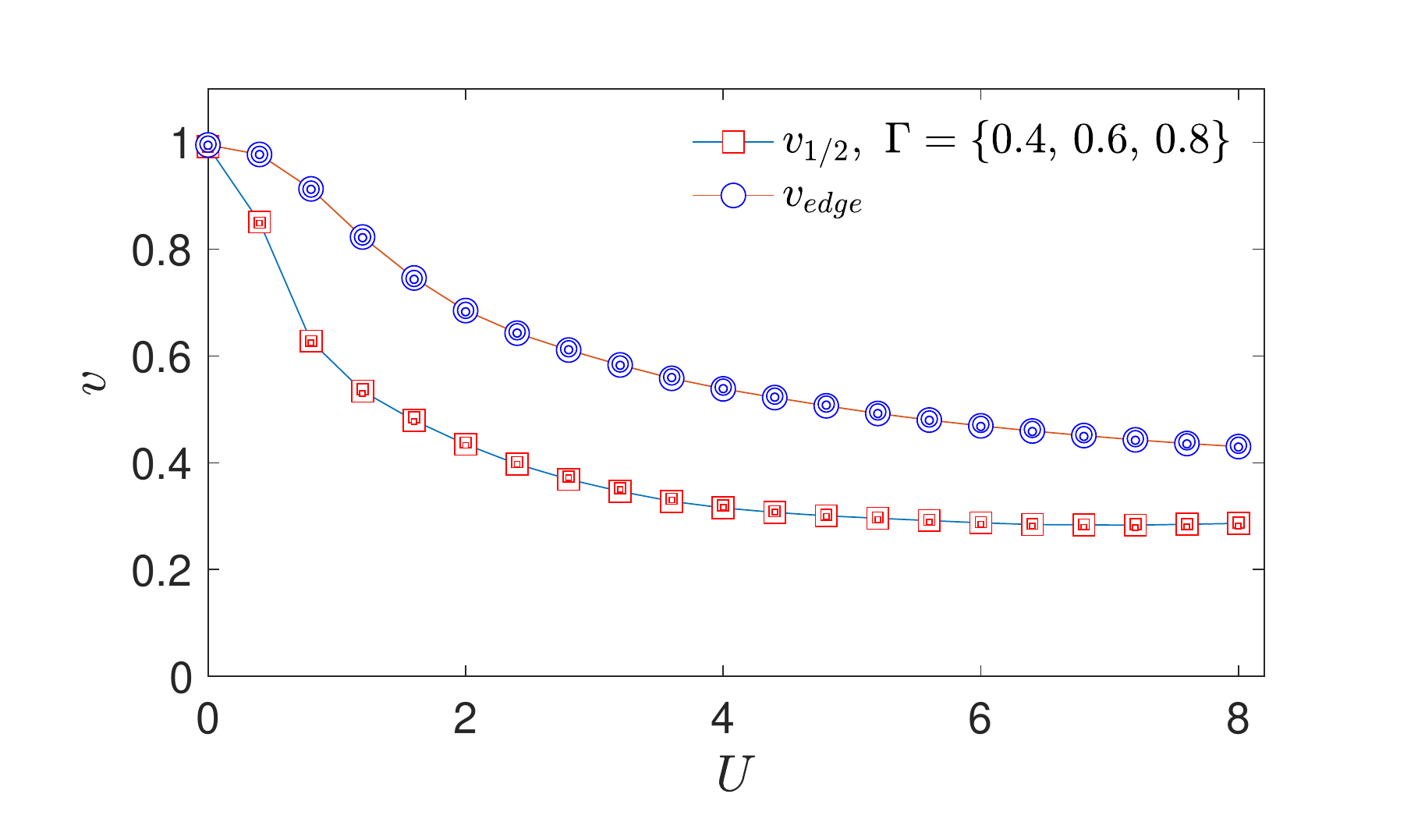}
\caption{ Front velocities $v_{1/2}$ and $v_\text{edge}$  as functions of $U$ for different decay rates, $\Gamma$, as extracted from the evolution of the current profile $\average{j(x,t)}$. 
For each velocity curve, different symbol sizes correspond to different values of $\Gamma$.
At $U=0$ we observe $v_\text{edge}=1$,  the  velocity of the fastest non-interacting electrons. Both velocities depend strongly 
on the interaction $U$ but show no dependence on $\Gamma$.}
\label{fig:front_velocity}
\end{center}
\end{figure}

\begin{figure}[b!]
\begin{center}
\includegraphics[width=0.95\columnwidth]{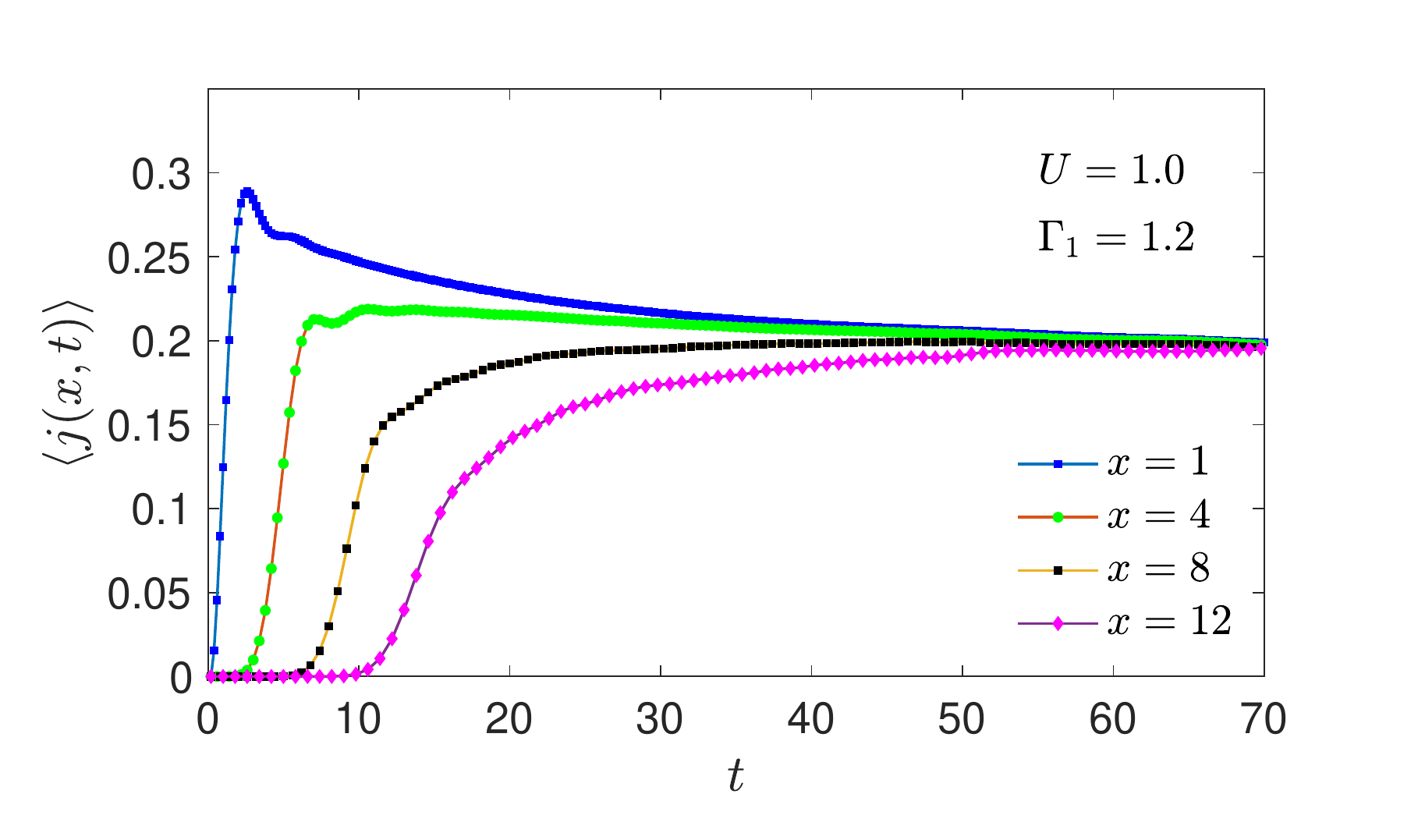}
\caption{ The time dependence of the average current $\langle j(x, t)\rangle$ for various sites along the chain of length $N=100$. 
}\label{fig:correlation_finite_U}
\end{center}
\end{figure}

This is quantitatively demonstrated in Fig.~\ref{fig:front_velocity}, where we display the front velocity for different values of 
$U$ and $\Gamma$.  The front velocity can be defined in several ways.   It can be defined as the velocity $v_\text{edge}$
of the  point, $x_{\rm edge}$, where the current raises beyond a pre-defined threshold. One can, however, also define it
 as the velocity of the point $x_{1/2}$,  which we denote as  $v_{1/2}$. As shown in Fig.~\ref{fig:front_velocity}, 
both velocities are reduced by interactions and smaller than the non-interacting velocity. At $U=0$ these two velocities coincide, 
but for finite $U$ they are clearly different. However, none of them seems to depend on the local disturbance, $\Gamma$. 
Also, both propagation velocities appear to scale to a finite   asymptotic value in the  $U\to 0$ limit.

The propagation of the current profile is accompanied by a depletion of particles for $x<x_{\rm edge}$, as demonstrated 
in the middle panels of Fig.~\ref{fig:curr_occup}. Quite strikingly, the slope of the $\langle n(x,t)\rangle $ curve is reduced with time, 
but the current remains finite. This clearly shows that, in spite of the interaction, no local equilibrium develops, even far from the front.
In local equilibrium, in distinction,
 the current should be induced by the gradient of the density, {$\langle j(x,t)\rangle\propto \partial_x\langle n(x,t)\rangle$}.

  An even more interesting front structure is observed in the entropy.
The  so-called  operator entanglement entropy,  $S^\text{op} (x,t)$,\cite{Prosen.2007} defined as the entanglement entropy of our vectorized state, 
provides  a certain measure of entanglement of the mixed state.  The evolution of this quantity is shown in the lower panels 
of Fig.~\ref{fig:curr_occup}. The operator entanglement  spreads also ballistically, however, rather surprisingly, it develops a two-step 
structure in the interacting case. The first step appears to move with a velocity $v_{\rm edge}$, together with the edge of the current profile. 
However, the true edge of the entropy profile seems to propagate faster than that, with a velocity, $v_0 \approx 1$, 
and penetrates deep into the 
$\langle n(x,t) \rangle \approx 1$ region, way before the depletion of particles reaches there.

The ballistic spread of the depletion region is natural in the non-interacting case, it is, however
quite surprising in the presence of interactions. A ballisticly spreading front implies namely a steady current 
density, $\lim_{t\to0}\langle j(x,t)\rangle = j_\infty  $ for all spatial coordinates $x$, and a corresponding 
linear increase of  the number of lost particles, $N_{\rm out}(t)$, i.e. particles that disappeared in the sink until time $t$. 
To obtain further confirmation of this ballistic behavior, we computed the current at various positions of the chain 
as a function of time.  This is shown in Fig. \ref{fig:correlation_finite_U} for ($U=1$). 
Within numerical accuracy, the current  indeed appears to approach asymptotically a steady state value, 
 independent of the location.

\begin{figure}[t!]
\begin{center}
\includegraphics[width=0.95\columnwidth]{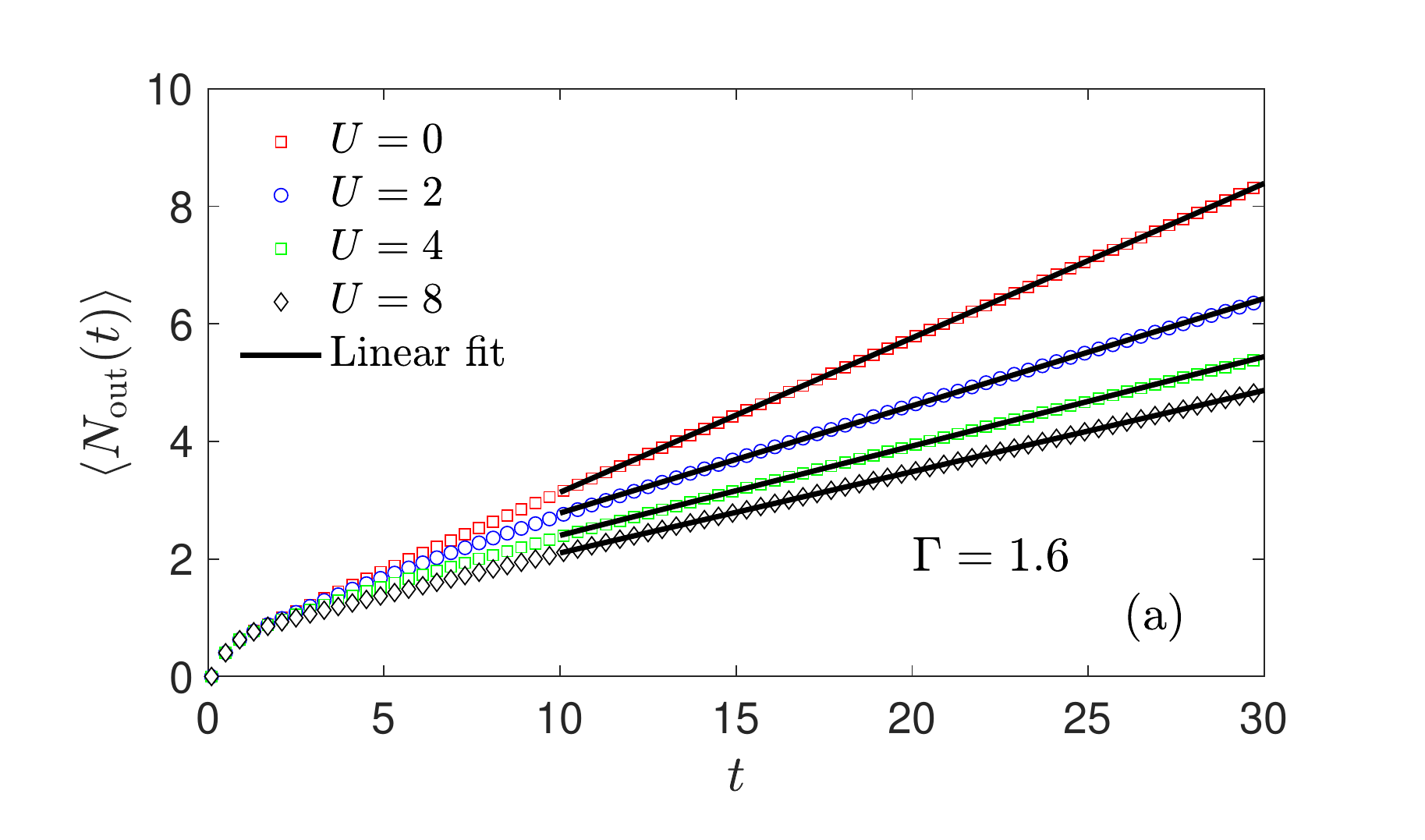}
\includegraphics[width=0.95\columnwidth]{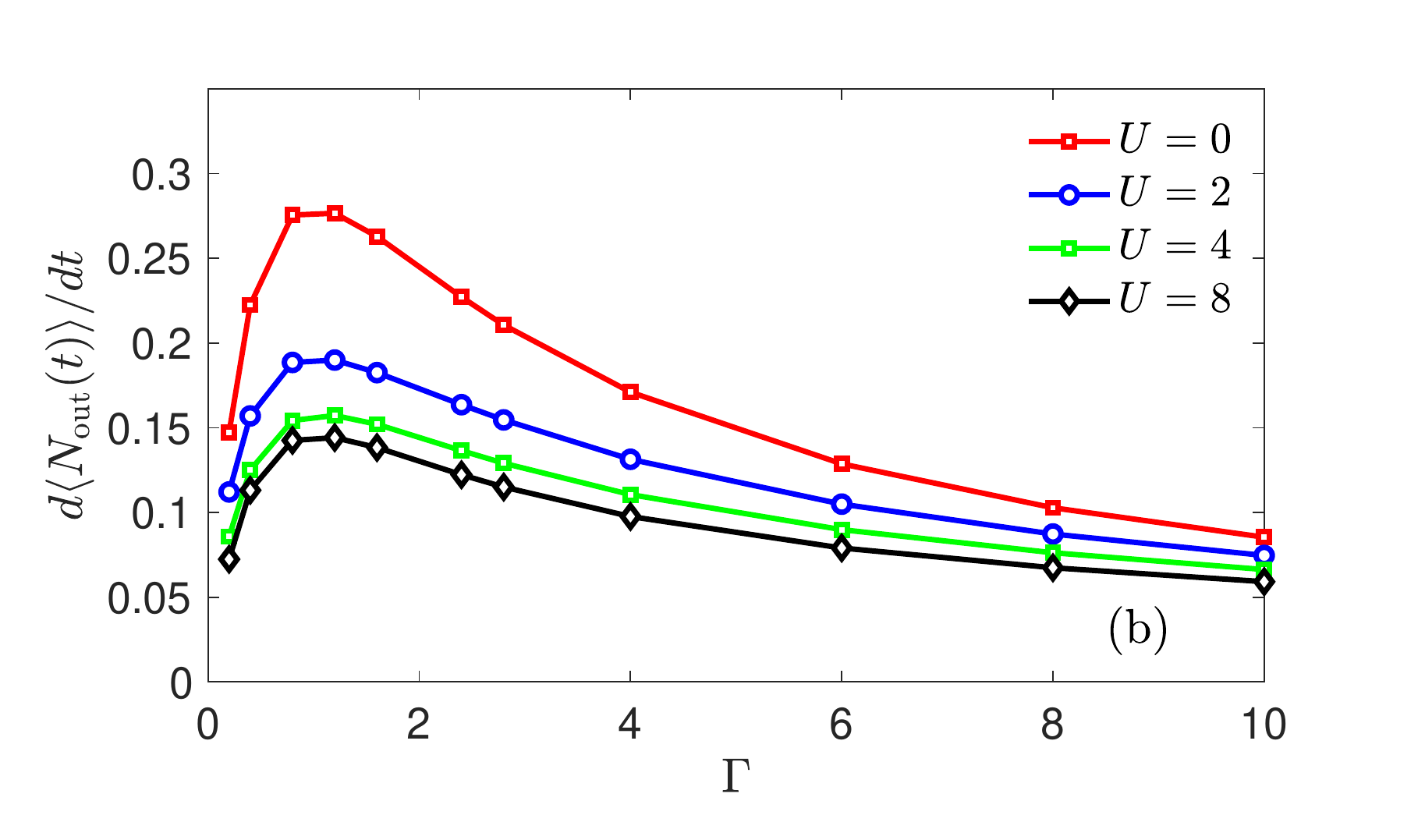}
\caption{ (a) Total particle loss as function of time for different interaction strengths. At large time
the total loss rate becomes asymptotically constant, indicating a linear increase of $\average{N_{\rm out}(t)}\sim \gamma\, t$ in time. (b) The total loss rate as function of $\Gamma$
in the quasi stationary regime $t\gg 1/\Gamma$, signaling the presence of the quantum Zeno effect.
 }\label{fig:loss}
\end{center}
\end{figure}

The total particle loss $N_{\rm out} (t)$ is shown in Fig.~\ref{fig:loss}(a) as a function of time. 
It is the largest  for $U=0$, and decreases with increasing $U$ such that at any given time $N_{\rm out}^{(U\ne 0)}(t)<N_{\rm tot}^{(U=0)}(t)$. 
For short times, $t\lesssim 1/\Gamma$, and large enough $\Gamma$,  the system is in a transition regime in which particles 
are lost exponentially at the first site, $N_{\rm out}(t\lesssim 1/\Gamma)\sim (1- e^{-\alpha t/\Gamma})$, where 
$\alpha$ depends on the initial filling but shows no or very weak dependence on the interaction.
At later times, the system reaches a quasi-stationary, non-equilibrium state
where the total loss increases linearly in time $N_{\rm out} (t\gg 1/\Gamma)\simeq \gamma\, t$ with a constant loss rate $\gamma = dN_{\rm out}/dt$ that depends on both $U$ and $\Gamma$.  The $\Gamma$ dependence is non-monotonic (see the inset in Fig.~\ref{fig:loss})
 which  can be viewed, similarly to the spinless case~\cite{Froml.2019,Wolff.2020},
as a  manifestation of the quantum Zeno effect~\cite{Barontini.2013, Froml.2020} (see Ref.~\onlinecite{Benenti} for a related effect in a boundary gain/loss driven chain).

\section{Conclusions}\label{sec:conclusions}
In this work we have developed a non-abelian time evolving block decimation (NA-TEBD) approach to investigate the 
dynamics of open systems possessing  non-Abelian symmetries. 
By  extending the notion of non-Abelian symmetries to the Lindbladian evolution, and organizing the superoperators as symmetry multiplets, 
we are able to construct an efficient NA-TEBD scheme that explicitly uses  abelian as well as  non-abelian symmetries of the Lindbladian. 

We applied this approach, to  the semi-infinite  SU(2) Hubbard model with loses at one end of the chain, having an $\textrm{SU(2)}\times \textrm{U(1)}$ 
symmetry.  In the non-interacting limit, $U=0$, we benchmarked our approach against the third quantization approach of Ref.~\onlinecite{Prosen.2008}, 
and investigated the time evolution of local density and the current operators as well as the operator entanglement following a dissipation quench
in a half-filled, infinite temperature state. In this case particles are depleted around the sink, and a propagating front appears, separating the 
depleted and occupied regions.

We investigated the structure of the propagating front in detail. Both in the non-interacting and in 
the interacting limits we find 
a ballistic front propagation. However, interactions reduce the light cone velocity substantially, $v_{\rm edge} < v_0 = J = 1$. 
In addition, interactions change, however, drastically the structure of the propagating front. 
For $U=0$, the front displays coherent fringes, and spreads as  $\sim t^{1/3}$,\cite{Eisler.2013} while for $U\ne0$ 
a ballistic $\sim t$ scaling of the whole current profile is observed, and the coherent fringes disappear. 

Corroborating these findings and the ballistic front propagation, we observe a saturation of the current to 
a   position independent asymptotic  value, $j_\infty (U,\Gamma)$, and a corresponding linear increase 
in the loss of particles, $\textrm{d}N/\textrm{d}t\to j_\infty$. 
The loss  rate is suppressed with increasing $U$, however, it exhibits a non-monotonous behavior 
as a function of the dissipation strength,  a clear manifestation of the quantum Zeno effect.

The   operator entanglement entropy,  $S^\text{op} (x,t)$,\cite{Prosen.2007} is also found 
to spread ballistically. 
However, the true edge of the entropy profile seems to propagate  with a velocity, $v_0  > v_{\rm edge}$, faster than the current profile, 
and penetrates deep into the $\langle n(x,t) \rangle \approx 1$ region, way before the depletion front reaches there.

The ballistic propagation we observe is somewhat surprising. It implies that the gradient of the 
density is \emph{unrelated} to the current density, implying, in turn, the absence of local equilibration.
Our simulations are consistent with this picture, while we cannot exclude, however, a slow, e.g.  logarithmic 
suppression of $\textrm{d}N/\textrm{d}t$. It is not clear, either, if the ballistic behavior observed has a relation to 
the integrability of the Hubbard model. 

These results demonstrate the power of this approach.
By  using explicitly all symmetries of the model, our approach makes it possible to study with high accuracy
the dynamics in open systems with relatively large local operator spaces, 
 even on platforms with limited computing power and memory, such as a regular desktop. 

\section*{Acknowledgments}

This research is supported by the National Research, Development and Innovation Office - NKFIH 
through research grants Nos.  K134983 and SNN139581, 
within the Quantum National Laboratory of Hungary program (Project No. 2017-1.2.1-NKP-2017-00001). 
M.A.W has also been supported by the \' UNKP-21-4-II New National Excellence Program of the National Research, 
Development and Innovation Office - NKFIH. C.P.M  acknowledges  
 support by the Ministry of Research, Innovation and Digitization, CNCS/CCCDI–UEFISCDI, under projects number PN-III-P4-ID-PCE-2020-0277.
O.L. acknowledges support from the Hans Fischer Senior Fellowship programme
funded by the Technical University of Munich -- Institute for Advanced Study and
from the Center for Scalable and Predictive methods for Excitation and Correlated phenomena (SPEC), funded as part of the Computational Chemical Sciences Program by the U.S.~Department of Energy (DOE), Office of Science, Office of Basic Energy Sciences, Division of Chemical Sciences, Geosciences, and Biosciences at Pacific Northwest National Laboratory.
T.P. acknowledges ERC Advanced grant 694544-OMNES and ARRS research program P1-0402.

\appendix

\section{Third quantization construction}\label{app:3rdQT}

In this section  we discuss in more detail the third quantization construction. Following Refs.~\cite{Prosen.2008,Kos.2017} we  introduce the Majorana basis,  and define the $4N$ operators $w_m$  as
\begin{gather}
w_{4x-3} = c_{x\uparrow}+c^\dagger_{x\uparrow}\phantom{aaaa}
w_{4x-2} = i(c_{x\uparrow}-c^\dagger_{x\uparrow})\nonumber\\
w_{4x-1} = c_{x\downarrow}+c^\dagger_{x\downarrow}\phantom{aaaa}
w_{4x} = i(c_{x\downarrow}-c^\dagger_{x\downarrow})\,.
\label{eq:Majorana_basis}
\end{gather}
By construction, they satisfy the anticommmutation relations $\{w_j, w_k\}=2\delta_{jk}$. In the Majorana basis\footnote{Any operator $A$
in the Majorana basis is denoted as \doubleunderline{A}, while in the original basis it is denoted as $ A$.} the matrix Hamiltonian \doubleunderline{H} corresponding to the hopping term in Eq. ~\eqref{eq:Hubbard} becomes
\begin{gather}
\doubleunderline{H} ={1\over 2} {H}\otimes \sigma_y. 
\end{gather}
The jump operators acting on the first site, are transformed to the Majorana basis as well,
and  take the general form $F_{1\sigma}\to \sum_{j}l_{\sigma, j}w_j$.  We can construct the associated 
Fock space $\cal K$ , i.e. the Liouville space, with dimension $2^{4N}$.  A typical orthonormal basis set consists of 
vectors of the form $\ketL {P_{\underline \alpha}}$, with $P_{\underline \alpha } =
 P_{\alpha_1\alpha_2\dots\alpha_{4N}}=w_1^{\alpha_1}w^{\alpha_2}\dots w_{4N}^{\alpha_{4N}}$, with $\alpha_j \in \{0,1\}$. 
The annihilation and the creation super-operators can be defined as 
\begin{gather}
\hat c_j \ketL{P_{\underline \alpha } } = \delta_{\alpha_j, 1}\ketL{w_j P_{\underline \alpha }}\,\phantom{aaa}
\hat c^\dagger_j \ketL{P_{\underline \alpha } } = \delta_{\alpha_j, 0}\ketL{w_j P_{\underline \alpha }}
\end{gather}
satisfying the cannonical anticommutation relations $\{ \hat c_j, \hat c^\dagger_j \} = \delta_{j,k}$. Keeping in mind that 
the dissipative part of the Lindbladian acts separetly in the even/odd sectors $\cal K = \cal K^{+}\oplus \cal K^{-}$,  we can restrict 
to one of these subspaces. Introducing the dissipative matrix $\doubleunderline M = \doubleunderline M^r +i \doubleunderline M^i$, 
where $\doubleunderline M^r$ and $\doubleunderline M^i$ are the 
real and imaginary parts of the matrix $M_{ij} = \sum_{\mu} l_{\mu, i} l^*_{\mu, j}$. The vectorized Lindblad equation becomes
\begin{gather}
i {d\over dt}\ketL{\rho(t)} = {\hat L }\,\ketL{\rho(t)}
\label{eq:Lindlbad_vec}
\end{gather}
with $ -i {\hat L} = -2 \hat c^{\dagger} \doubleunderline{X}^T \hat c  +4 i \hat c^{\dagger} {\doubleunderline M}^i \hat c$. Here  
$\doubleunderline {X}^T  = 2 i \doubleunderline H +2 \doubleunderline M^r$. In this way the vectorized  Lindblad equation for the density matrix 
is mapped to a regular Schr\" odinger equation in the Fock super-operators space. To compute the expectation values and the correlators, 
one defines the left Liouvillian vacuum, $\bra{\mathds{1}} = \bra{P_{0,0,\dots 0}}$, in terms of which any two-point correlator (covariance) can be written
as $\average{w_j w_k} = \bra{\mathds{1}} \hat c_j \hat c_k \ket {\rho}+\delta_{jk}$. The covariance matrix $ C_{ij} \equiv \average{w_i w_j}$ 
satisfies the associated equation 
\begin{gather}
{d\over dt}\doubleunderline{C}(t) = -2 \doubleunderline{X}^T \doubleunderline{C} (t) - 2\doubleunderline{C}(t)\doubleunderline{X} - 8 i \doubleunderline{M}^{i},
\end{gather}
which can be solved by standard methods, starting from some initial condition $\doubleunderline{C}(0)$. Physical observables can be computed 
from $C_{ij} \equiv \average{w_i w_j}$. The average occupation 
number along the Hubbard chain can be obtained, e.g., as 
\begin{gather}
\average{n_j(t)} = 1- {i \over 2}\left (C_{4j-3, 4j-2}(t)+ C_{4j-1, 4j}(t) \right ),
\end{gather}
and other  operators defined on the chain can be constructed in a similar manner. 

\bibliography{references}
\end{document}